\documentclass[pdflatex,sn-mathphys-num]{sn-jnl}% Math and Physical Sciences Numbered Reference Style
\usepackage{graphicx}%
\usepackage{subcaption}
\usepackage{multirow}%
\usepackage{amsmath,amssymb,amsfonts}%
\usepackage{amsthm}%
\usepackage{mathrsfs}%
\usepackage[title]{appendix}%
\usepackage{xcolor}%
\usepackage{textcomp}%
\usepackage{manyfoot}%
\usepackage{booktabs}%
\usepackage{algorithm}%
\usepackage{algorithmicx}%
\usepackage{algpseudocode}%
\usepackage{listings}%
\usepackage{mdframed}%
\usepackage{xcolor}

\theoremstyle{thmstyleone}%
\theoremstyle{thmstyletwo}%

\theoremstyle{thmstylethree}%

\begin{document}

\title[Executable verification]{Executable verification through formalized expert reasoning in astronomical spectroscopy}

\author[1,2]{\fnm{Haosong} \sur{Wang}}
\author*[3]{\fnm{Ting} \sur{Tan}}\email{ting.tan@cea.fr}
\author[1]{\fnm{Ji} \sur{Yao}}
\author[1]{\fnm{Jiajun} \sur{Zhang}}
\author[1]{\fnm{Qian} \sur{Zheng}}
\author[3]{\fnm{Christophe} \sur{Yeche}}
\author[4]{\fnm{Jean-Paul} \sur{Kneib}}
\author*[1]{\fnm{Huanyuan} \sur{Shan}}\email{hyshan@shao.ac.cn}

\affil[1]{\orgdiv{Shanghai Astronomical Observatory}, \orgname{Chinese Academy of Sciences}, \orgaddress{Shanghai 200030, P. R. China}}
\affil[2]{\orgname{University of Chinese Academy of Sciences}, \orgaddress{Beijing 100049, P. R. China}}
\affil[3]{\orgname{CEA, IRFU, Universit\'e Paris-Saclay}, \orgaddress{F-91191 Gif-sur-Yvette, France}}
\affil[4]{\orgname{Laboratoire d'Astrophysique, Ecole Polytechnique F\'ed\'erale de Lausanne (EPFL), Observatoire de Sauverny}, \orgaddress{CH-1290 Versoix, Switzerland}}

\abstract{Artificial intelligence has reshaped scientific prediction, but scientific verification remains a human bottleneck. Automated systems can map observations to labels, parameters or hypotheses, yet scientific conclusions require evidence, must satisfy physical consistency, and need explicit testing of alternatives before a decision is made. Here we introduce \textsc{FORMA} (Formalized Observational Reasoning with Auditable Decisions), an executable verification protocol that reconstructs expert reasoning into a workflow: it extracts evidence, generates hypotheses under physical constraints, tests alternatives, and performs auditable consistency checks. Unlike prediction or post-hoc interpretability, executable verification records and tests the evidential path leading to a decision. Astronomical spectroscopy provides a natural testbed, because ambiguous survey spectra are still adjudicated by expert visual inspection. Applied to the Dark Energy Spectroscopic Instrument (DESI) visual inspection catalogue, \textsc{FORMA} combines template-fitting candidate redshifts, spectral evidence extraction and physical audit into an auditable credibility score. A medium-or-higher credibility threshold identifies $331$ definite predictions with $95.5\%$ binary agreement with expert-adjudicated classes, while increasing credibility is associated with improved redshift consistency and higher classification reliability. These results show that automated inference can be coupled to explicit verification, allowing candidate outputs to be evaluated before they enter scientific use.}


\maketitle

Artificial intelligence has reshaped the generation of scientific prediction, but it has not yet scaled scientific verification, which remains largely dependent on human expertise. In observational and experimental sciences, expert validation does more than assign labels to data: it transforms instrumental evidence into physical hypotheses, verifies them against domain principles and alternatives, and decides whether a conclusion should be accepted, rejected or deferred. This evidential process is central to scientific inference, because reliable conclusions require more than statistical plausibility; they require falsifiability, reproducibility and traceable evidence~\citep{Rudin2019Interpretable}. This verification requirement presents an immediate data-processing bottleneck for next-generation, large-scale surveys.

Astronomical spectroscopy provides a particularly well-defined testbed for addressing this challenge since expert validation already follows a structured protocol grounded in astrophysical first principles. Machine-learning and template-fitting methods have substantially advanced spectroscopic classification and redshift estimation, including neural-network classifiers, survey-optimized redshift-fitting pipelines and recent language-model-based approaches to spectral interpretation~\citep{Busca2018QuasarNet,redrock2024,ramachandra2025teachingllmsspeakspectroscopy}. More broadly, large language models, vision–language systems and autonomous agents are increasingly being applied across astronomical and physical sciences, including observatory operations, data retrieval, transient interpretation, multi-band galaxy analysis, gravitational-wave search, solar-flare forecasting, radio astronomy, lensing inference and general scientific computing ~\citep{shao2025jwflareaccuratesolarflare, wang2025starwhisper,kostunin2025enhancingdevelopmentcherenkovtelescope,tamhane2025naturallanguageinterfaceefficient,wang2025automated,miao2025physmasterbuildingautonomousai,drozdova2025radioastronomyeravisionlanguage,ting2025egentautonomousagentequivalent,stoppa2025textual,sun2025mephistoselfimprovinglargelanguage,parker2025aion1omnimodalfoundationmodel,feng2026lensagentselfevolvingagent}. These systems expand predictive, interpretive or operational capability. Expert review, however, remains the definitive mechanism for resolving ambiguous classifications and redshift determinations in modern survey pipelines~\citep{Lan2023DESIVI,Dawson2016eBOSS,alexander23}, because it tests physical consistency, alternative explanations and evidential sufficiency.

Here we show that expert reasoning, composed of hypothesis testing, falsification and physical consistency, can be formalized as an executable computational protocol. We introduce \textsc{FORMA} (Formalized Observational Reasoning with Auditable Decisions), a protocol that reconstructs astronomical spectroscopy expert validation into a structured process of evidence extraction, physical hypothesis generation, systematic evaluation of alternative explanations and independent consistency checks. By embedding domain rules as rigid constraints and preserving a complete evidential record for every inference, \textsc{FORMA} converts automated inference from an opaque prediction process into a transparent verification workflow. In spectroscopic implementation, \textsc{FORMA} makes use of template-fitting engines to generate candidate redshift hypotheses that are passed to the verification workflow as proposed interpretations, while \textsc{FORMA} evaluates whether each interpretation is physically and evidentially admissible. \textsc{Redrock}\footnote{ https://github.com/desihub/redrock} is used as the DESI-optimized example of such a template-fitting engine, but the framework is not specific to \textsc{Redrock} and can in principle operate with any engine that supplies candidate redshift solutions.

\section*{1. A protocol for executable scientific verification}
\textsc{FORMA} implements scientific verification in astronomical spectroscopy as an executable protocol, reconstructing expert spectral review into a multi-agent workflow (Fig.~\ref{fig1}). The protocol is organized around five linked requirements: observational evidence represented in physically meaningful form, with uncertainty and instrumental context; hypothesis generation constrained by domain knowledge; explicit testing of alternative explanations to avoid accepting statistically plausible but physically unsupported inferences; reproducible physical reasoning checks that are separable from language-model reasoning, and a recorded evidential path leading to a credibility score. These five requirements collectively ensure that every decision is traceable, physically interpretable, falsifiable, and reproducible, while providing a clear gating mechanism for safe large-scale deployment.

Fig.~\ref{fig1} separates this protocol from its spectroscopic implementation. The conceptual framework is shown in panels a and b, where expert visual reasoning is converted into machine-readable evidence; panel c shows how this evidence is passed through constrained hypothesis generation, iterative audit, synthesis and report construction. The workflow begins with a Visual Interpreter that converts an input spectrum into an evidence representation mirroring expert inspection, including continuum structure, troughs, peaks, candidate features and redshift-relevant patterns. A Hypothesis Analyst then constructs physically allowed interpretations by combining redshift–wavelength relations, diagnostic line associations and domain knowledge. Candidate redshifts from template-fitting tools can enter this stage as external hypotheses, but they are not treated as accepted conclusions; they are subjected to the same feature-level, line-identification and physical-consistency audits as hypotheses generated internally by \textsc{FORMA}. Candidate interpretations are evaluated through hypothesis synthesis and audit steps, which test whether the proposed interpretation is supported by the observed evidence, whether expected companion features are present or absent, whether alternative explanations remain viable and whether deterministic consistency checks are satisfied. Failed or incomplete hypotheses could re-enter the workflow for further inspection, rather than being forced into a final label. A Report Writer consolidates the accepted interpretation, rejected alternatives, unresolved concerns and the \textsc{FORMA} credibility score into an auditable scientific report. External tools and a knowledge base provide reproducible numerical validation and domain constraints throughout the workflow.

We emphasize that this work presents a general verification protocol, with FORMA as its spectroscopic reference implementation, rather than a standalone classification algorithm. \textsc{FORMA} is one implementation of the protocol in a domain where expert validation is well defined and survey-scale bottlenecks are acute. In this formulation, scientific verification becomes executable because each step of the evidential process can be inspected, reproduced and challenged. The resulting system functions as an auditable verification layer: it evaluates whether a scientific inference is sufficiently supported by available evidence, instead of generating predictions that must await human review.

\section*{2. Executable verification on the DESI expert-review catalogue}

To evaluate whether scientific verification can be rendered executable across realistic survey conditions, we applied \textsc{FORMA} to 
%\textcolor{orange}{an 1,149-sample evaluation set constructed from} 
the DESI Early Data Release (EDR) visual-inspection (VI) catalogue\footnote{https://data.desi.lbl.gov/doc/releases/edr/vac/vi}. The analysis includes spectra across the available signal-to-noise (SNR) range, including cases in which expert review is normally required because the evidence is incomplete, ambiguous or potentially misleading. This setting directly tests the intended role of executable verification: its purpose is not to assign a label to every spectrum, but to check whether the evidence supports acceptance—or whether the case should be deferred or rejected.

All spectra were analysed using the complete \textsc{FORMA} verification workflow in a zero-shot setting without dataset-specific training. We evaluate FORMA with DeepSeek-V4-Pro \cite{deepseekv4} as the base LLM, accessed via API. The \textsc{FORMA} credibility score is assigned after evidence extraction, physics-constrained hypothesis generation, alternative testing and consistency checks. This score is distinct from conventional LLM self-confidence metrics. Instead, it quantifies the sufficiency of supporting evidence as reconstructed by the agent workflow: high credibility corresponds to evidence robust enough for autonomous verification, while low-credibility cases are flagged for manual audit or expert review.

For each spectrum, \textsc{FORMA} was supplied with the observed spectral data and candidate redshift hypotheses from a template-fitting engine. In the DESI application, \textsc{Redrock} provides the survey-optimized template-fitting solutions and is therefore used to generate potential redshift hypotheses. \textsc{FORMA} uses these candidate redshifts only as hypotheses to be tested; external pipeline classifications and human visual-inspection labels are used only for evaluation and are not supplied as accepted answers during inference. The \textsc{FORMA} credibility score defines a practical triage boundary across the full evaluation sample (Fig.~\ref{fig2}a). Moving from the low-or-higher to the medium-or-higher threshold reduces the definite-classification pool from $70.4\%$ to $28.9\%$ of spectra, while increasing agreement with expert-adjudicated classes from $90.0\%$ to $95.5\%$. A higher threshold yields a smaller but purer subset, with $18$ high-credibility cases and $100.0\%$ classification accuracy in this evaluated sample; however, this tier is too small to serve as the main operational catalogue boundary. We therefore use the medium-or-higher credibility threshold to evaluate the definite-prediction candidate pool.

At this threshold, \textsc{FORMA} produces $331$ definite predictions (Fig.~\ref{fig2}b). Outputs labelled as unknown are treated as deferred cases rather than forced classifications and are excluded from the definite-prediction matrix. The remaining candidates reproduce expert adjudication with substantial fidelity: \textsc{FORMA} agrees with expert quasar (QSO) classifications for $100$ of $103$ cases and with expert galaxy classifications for $216$ of $228$ cases, corresponding to an overall binary agreement of $95.5\%$. The per-class agreement rates are $97.1\%$ for QSOs, $93.8\%$ for LRGs, $94.5\%$ for ELGs and $95.0\%$ for BGSs. Off-diagonal candidate disagreements are not treated as silently accepted catalogue entries; they are cases that require audit filtering or expert review.

The \textsc{FORMA} credibility score also correlates with the reliability of continuous physical quantities (Fig.~\ref{fig2}c). Redshift residuals are broad for low-credibility spectra and become progressively more concentrated at medium and high credibility. The adopted agreement criterion $\Delta z\le 0.01$ provides a reference tolerance for visual-inspection-level consistency. This behavior indicates that the credibility score reflects evidential support for both classification and redshift inference.

Finally, the SNR dependence of low-credibility spectra shows that \textsc{FORMA} credibility is not simply a proxy for raw data quality (Fig.~\ref{fig2}d). We assign a \texttt{SNR\_sum}, which is the summation of \texttt{MEDIAN\_COADD\_SNR\_B/R/Z}, to each spectrum, as an indicator of overall SNR. Even within the low-credibility tier, classification accuracy and redshift residuals vary with spectral SNR ratio. This demonstrates that low credibility arises from insufficient or ambiguous evidence as evaluated by the verification pipeline, and is not simply a proxy for low spectral SNR. Together, these classification, redshift and SNR-dependent tests show that \textsc{FORMA} credibility tracks the sufficiency of the supporting evidence, not simply the apparent quality of the spectrum or the confidence of an underlying model.

Visual inspection provides the expert-adjudicated reference used here, but its quality flag was not designed to be an object-level measure of evidential sufficiency. \textsc{FORMA} instead assigns credibility from the explicit availability, consistency and uniqueness of the spectral evidence, making low-credibility cases auditable rather than silently accepted. We discuss this distinction in Appendix~\ref{app:feature-score}.

\section*{3. From executable verification to scientific conclusions}

The credibility-gated triage shown in Fig.~\ref{fig2} addresses verification at the catalogue scale, determining which entries can be safely adopted for scientific analysis. Beyond bulk catalogue processing, verification also operates on individual sources by testing the robustness of interpretations and recovering physically meaningful classifications missed by standard pipelines. We illustrate both functions through two representative cases.

The first role of verification is to prevent statistically plausible but physically inconsistent interpretations from being accepted as scientific conclusions. In the workflow considered here, \textsc{FORMA} does not replace specialized redshift-fitting engines. Instead, it evaluates whether candidate interpretations are physically and evidentially admissible. This distinction matters because verification is not a second classification step; it tests whether a proposed physical interpretation is sufficiently supported, should be rejected, or must be deferred.

To test whether \textsc{FORMA} rejects interpretations whose supporting evidence has been weakened or made physically inconsistent, we constructed counterfactual evidence-ablation tests (Fig.~\ref{fig3}a). Starting from an observed spectrum with diagnostic QSO-like features, we modified the evidence-bearing regions, as well as corresponding catalogs, while leaving the rest of the spectrum unchanged (see Appendix~\ref{app:counterfactual}). In one counterfactual, the Ly$\alpha$ region is removed; in another, broad emission lines are narrowed to line widths of $900$--$1000~\rm km/s$. These modifications preserve a spectrum-like input but remove or alter the physical evidence required for a broad-line QSO interpretation. \textsc{FORMA} assigns reduced credibility and rejects or defers the modified interpretation; it does not treat it as an accepted conclusion. This test confirms that \textsc{FORMA} validates interpretations based on the physical integrity of their supporting evidence, rather than merely matching templates to assign labels.

Executable verification also helps recover physically meaningful interpretations that may be suppressed by standard catalogue labels or ordinary galaxy interpretations (Fig.~\ref{fig3}b). In the representative low-luminosity AGN case, the object is treated as a galaxy in the standard classification layer, whereas expert visual inspection identifies a host-dominated low-luminosity AGN candidate. \textsc{FORMA} supports the expert-VI interpretation by detecting high-ionization and AGN-relevant diagnostics, including [Ne~{\sc v}], strong [O~{\sc iii}] and C~{\sc iii}], and flags the object as a verification-supported AGN candidate. This case should not be presented as a confirmed discovery; rather, Fig.~\ref{fig3}b shows that \textsc{FORMA} can make explicit the evidential basis for an expert-supported physical interpretation and route it to audit or follow-up when needed.

Together, Fig.~\ref{fig3}a and Fig.~\ref{fig3}b illustrate complementary functions of executable verification. The counterfactual evidence-ablation tests show that \textsc{FORMA} can reject or defer an interpretation when its diagnostic evidence is removed or made physically inconsistent, while the low-luminosity AGN case shows that \textsc{FORMA} can surface an expert-supported physical interpretation that may be suppressed by standard classification labels. In both cases, the verification layer filters physical interpretations without replacing specialized inference engines. 

\section*{4. Discussion and conclusions}

This work reframes scientific verification as an explicit scientific protocol. Predictions may be generated by theories, simulations, statistical models or artificial intelligence, but scientific conclusions are accepted only after they have been tested against evidence, alternatives and domain constraints. As scientific datasets grow, this verification step has become a fundamental scalability bottleneck. The results presented here show that verification need not remain an exclusively human activity: it can be formalized as an executable protocol.

The central contribution of this work is the formulation and demonstration of such a protocol. \textsc{FORMA} provides a reference implementation of this protocol in astronomical spectroscopy, where expert validation is well defined and survey-scale bottlenecks are acute. In this implementation, evidence extraction, constrained hypothesis generation, alternative testing, independent consistency checks and credibility-gated decisions are assembled into a closed verification loop. Applied to the DESI EDR expert-review catalogue across all SNR regimes, this loop identifies ambiguous or unsupported cases, improves classification reliability at higher credibility thresholds and prevents physically inconsistent interpretations from being propagated without warning. Medium-or-higher credibility cases show strong agreement with expert classifications and improved redshift consistency, while the SNR-dependent diagnostics show that low credibility is not simply a proxy for low SNR. Failures and ambiguous cases are therefore not hidden as silent catalogue entries, but represented explicitly as cases to audit, reject or defer. Appendix~\ref{app:feature-score} further contrasts this evidence-based credibility with conventional human visual-inspection quality flags, emphasizing that the latter are indispensable for expert adjudication but do not by themselves encode the full evidential path required for reproducible verification.

This approach distinguishes executable verification from both predictive modelling and post-hoc interpretability. Predictive models are optimized to generate outputs from input data, while interpretability methods seek to explain those outputs retroactively. By contrast, executable verification integrates validation into the inference workflow: the full evidential chain leading to acceptance, rejection, or deferral is built and retained as an explicit computational product. In the DESI expert-review catalogue, this appears as a transition from human-centred visual adjudication to a \textsc{FORMA} credibility score. The \textsc{FORMA} credibility score is therefore not ordinary LLM self-confidence, but an agent-derived reconstruction of the evidential sufficiency judgment normally supplied by expert visual inspection. The robustness of the evidence-ablation test is confirmed by $100$ independent runs under varying knowledge-base conditions (Supplementary Table.~B1), and the insufficiency of simple feature counting is demonstrated in Supplementary Fig.~\ref{fig:app-score}.

The protocol is not intrinsically tied to spectroscopy. Within astronomy, the same structure could be instantiated with different evidence and consistency checks: light curves and host context for transients, transit shape and centroid tests for exoplanets, or image multiplicity and lens-equation closure for strong lenses. More broadly, what transfers is the evidential protocol: automated candidate interpretations are tested against physical alternatives before they are accepted, rejected or deferred.

The present implementation has limitations. \textsc{FORMA} does not replace dedicated high-precision inference engines such as template-fitting redshift pipelines; rather, it evaluates whether their outputs, or its own hypotheses, are physically and evidentially admissible. Its redshift estimates are therefore complementary to optimized fitting methods, and future versions should integrate high-precision template hypotheses within the verification loop. Similarly, candidate physical interpretations such as host-dominated low-luminosity AGNs should be treated as verification-supported hypotheses unless independently confirmed by additional diagnostics or follow-up data. These limitations define the boundary between autonomous acceptance and required expert judgment. We note that \textsc{FORMA} does not rely on vision-language models to interpret rendered spectral images. All primary evidence extraction operates directly on numerical flux arrays via deterministic feature detection; language models are used only for physical reasoning and audit, not for raw spectral reading. This design avoids the resolution and blending artifacts to which generic VLMs are prone. 
A recent systematic evaluation of astronomical VLMs found that performance depends strongly on data representation and physical grounding, and that correct labels can still be accompanied by physically incomplete reasoning~\cite{ren2026}; this supports \textsc{FORMA}'s use of models within a tool-supported verification protocol rather than as a substitute for evidence extraction and physical audit.

Consequently, executable verification suggests a new layer of scientific infrastructure for large surveys. Future facilities such as the Vera C. Rubin Observatory \cite{Ivezic2019LSST}, Square
Kilometre Array Observatory (SKAO) \cite{SKAO2020}, CMB-S4 \cite{cmbs4}, {\it Euclid} \cite{euclid2011, euclid2025}, and Chinese
Space-station Survey Telescope (CSST) \cite{csst2026}, alongside spectroscopic surveys including 4-metre Multi-Object Spectroscopic Telescope (4MOST) \cite{4most}, Prime Focus Spectrograph (PFS) \cite{PFS2022, PFS2024}, MUltiplexed Survey Telescope (MUST) \cite{MUST_zhao, MUST_Cai_2025}, Jiaotong
University Spectroscopic Telescope (JUST) \cite{JUST2024} and DESI II, will require not only predictive engines that generate candidate inferences, but also auditable verification protocols that preserve falsifiability, reproducibility and evidential traceability. In \textsc{FORMA}, each accepted, rejected or deferred inference is accompanied by an explicit record of the supporting observations, rejected alternatives, consistency checks and provenance. When accumulated in versioned knowledge repositories, such records can preserve operational expertise that is otherwise only partially captured in publications, technical notes or individual experience. This makes executable verification useful not only for individual decisions, but also for knowledge transfer, reproducible review and continual refinement across long-lived collaborations. In this sense, executable verification provides a mechanism for preserving scientific judgment alongside scientific data.

\textbf{Author contributions:} HSW: designed the method, analyzed the data and results, drafted the manuscript; TT: conceived the idea, designed the method, analyzed the data and results, drafted the manuscript; JY: analyzed the results, edited the manuscript; JJZ: analyzed the results, edited the manuscript; QZ: analyzed the results, edited the manuscript; CY: edited the manuscript; JPK: edited the manuscript; HYS: conceived the idea, designed the method, analyzed the results, drafted the manuscript.

\section*{Code availability}
Github link of \textsc{FORMA} Agent: https://github.com/SpecSurvey-In-AI-era/FORMA\\

\clearpage

% ============================================================
% FIGURES
% ============================================================

\begin{figure*}[p]
\centering
\includegraphics[width=1.\textwidth]{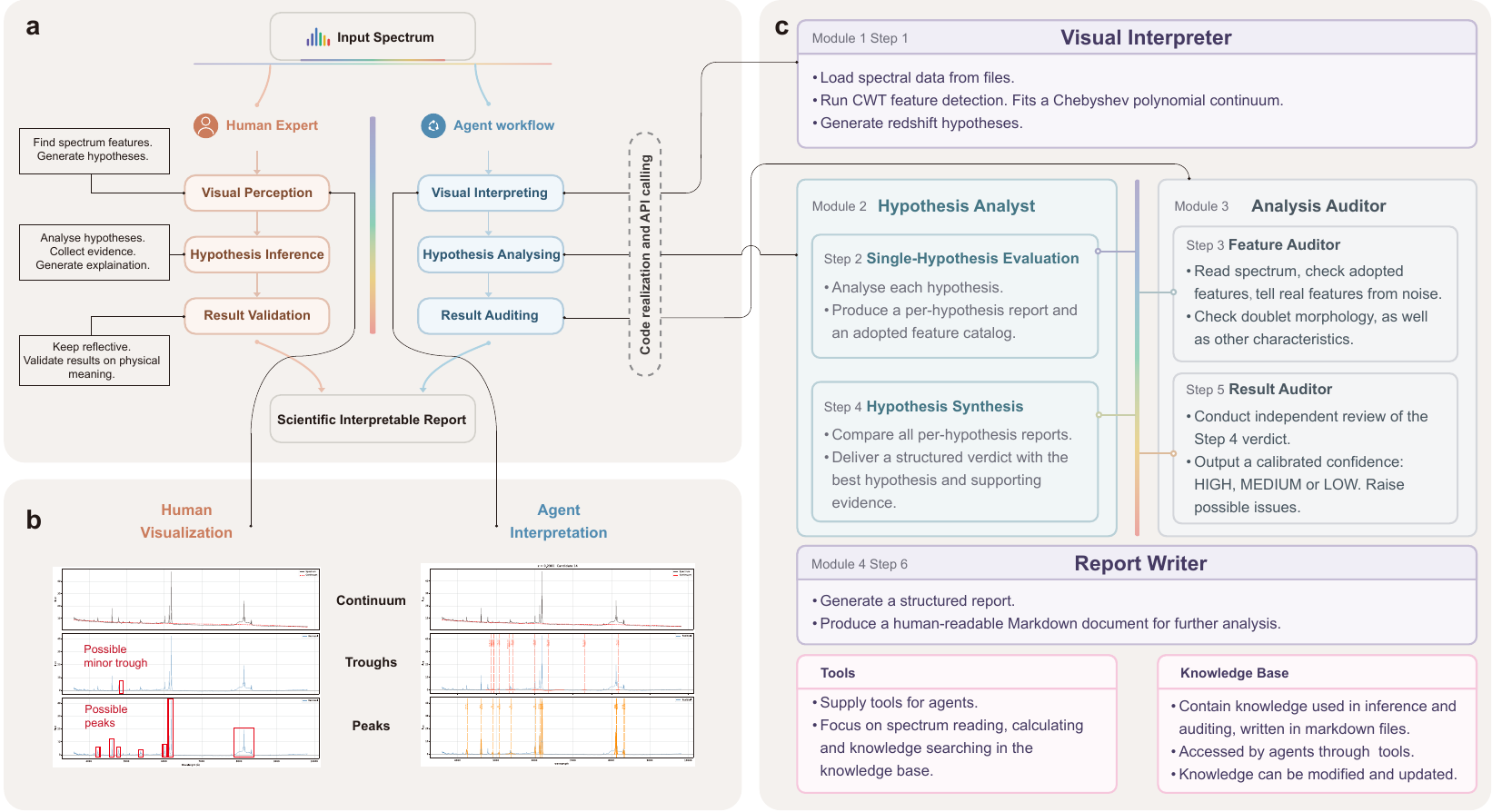}
\caption{\textbf{Computational architecture of executable scientific verification.} \textbf{(a)} Scientific verification is formulated as a transformation from human-centred expert validation to agent-centred executable verification. Human experts convert visual perception into hypothesis inference and result validation; FORMA mirrors this structure through visual interpretation, hypothesis analysis and result auditing, producing an auditable scientific report. \textbf{(b)} Example of the evidence representation used by human and agent workflows. Human inspection identifies continuum structure, troughs and peaks, while the agent workflow reconstructs analogous spectral evidence as computational features for downstream hypothesis testing. \textbf{(c)} Modular implementation of FORMA in astronomical spectroscopy. The Visual Interpreter extracts spectral evidence; the Hypothesis Analyst constructs physically constrained interpretations; the Analysis Auditor tests candidate hypotheses and consistency constraints; and the Report Writer records the accepted interpretation, rejected alternatives and FORMA credibility score. External tools and a knowledge base support reproducible validation and domain-constrained reasoning.}
\label{fig1}
\end{figure*}

\begin{figure*}[p]
\centering
    \includegraphics[width=0.99\linewidth]{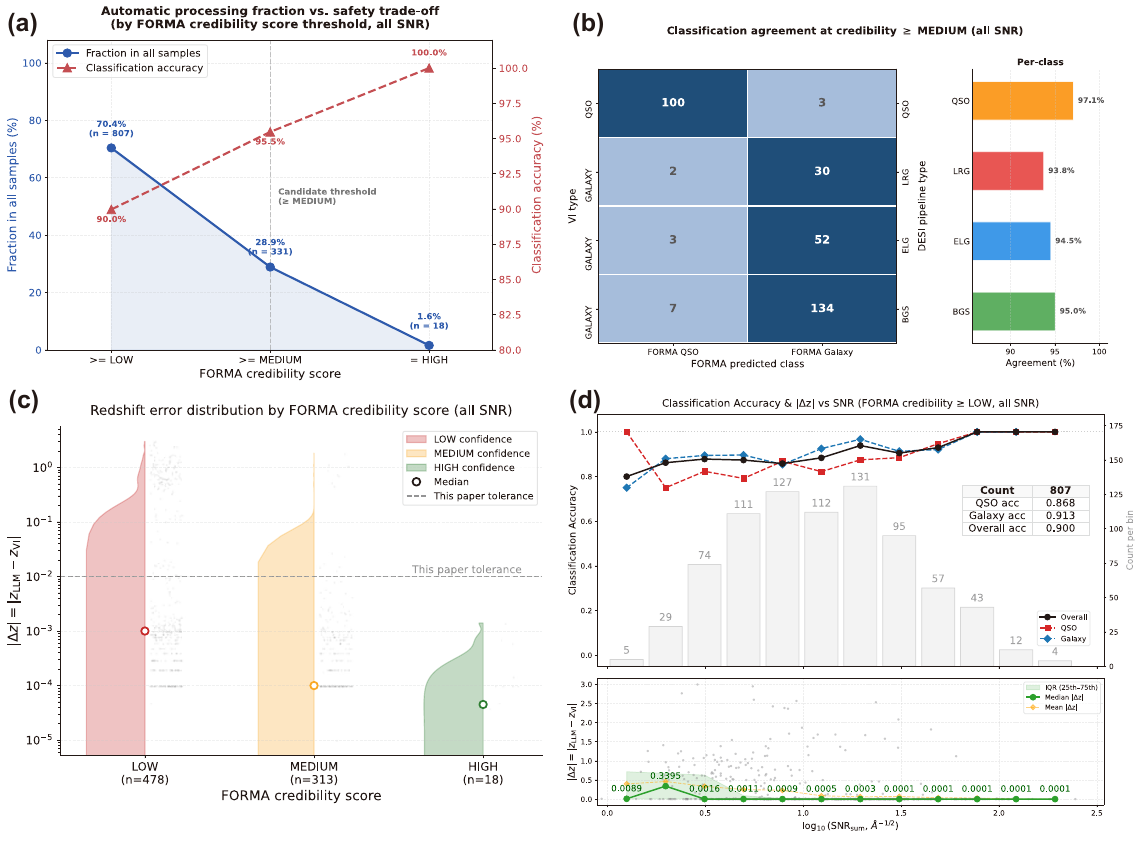}
    \caption{\textbf{FORMA credibility scoring across the DESI expert-review catalogue.} The analysis uses the DESI EDR expert-review catalogue without an additional SNR cut. \textbf{(a)} Fraction of spectra routed toward automatic verification and classification accuracy, measured against expert-adjudicated classes, as functions of the FORMA credibility-score threshold. Medium-or-higher credibility provides the operational candidate threshold used in panel~b. \textbf{(b)} Classification agreement for medium-or-higher credibility candidates. The matrix shows expert-adjudicated type versus FORMA predicted class, and the right panel shows per-class agreement rates for QSOs, LRGs, ELGs and BGSs. Unknown outputs are treated as deferred cases and are excluded from the definite-prediction matrix. \textbf{(c)} Redshift residual distributions as a function of FORMA credibility tier. Higher credibility corresponds to more concentrated residuals, with the dashed line marking the reference tolerance $|\Delta z|\le 0.01$. (d) Classification accuracy and redshift residuals as functions of SNR ratio within the low-credibility tier. This panel shows that low credibility is not equivalent to low SNR alone, but reflects the evidential sufficiency assessed by the verification workflow.}
    \label{fig2}
\end{figure*}

\begin{figure*}[p]
\centering
    \includegraphics[width=0.99\linewidth]{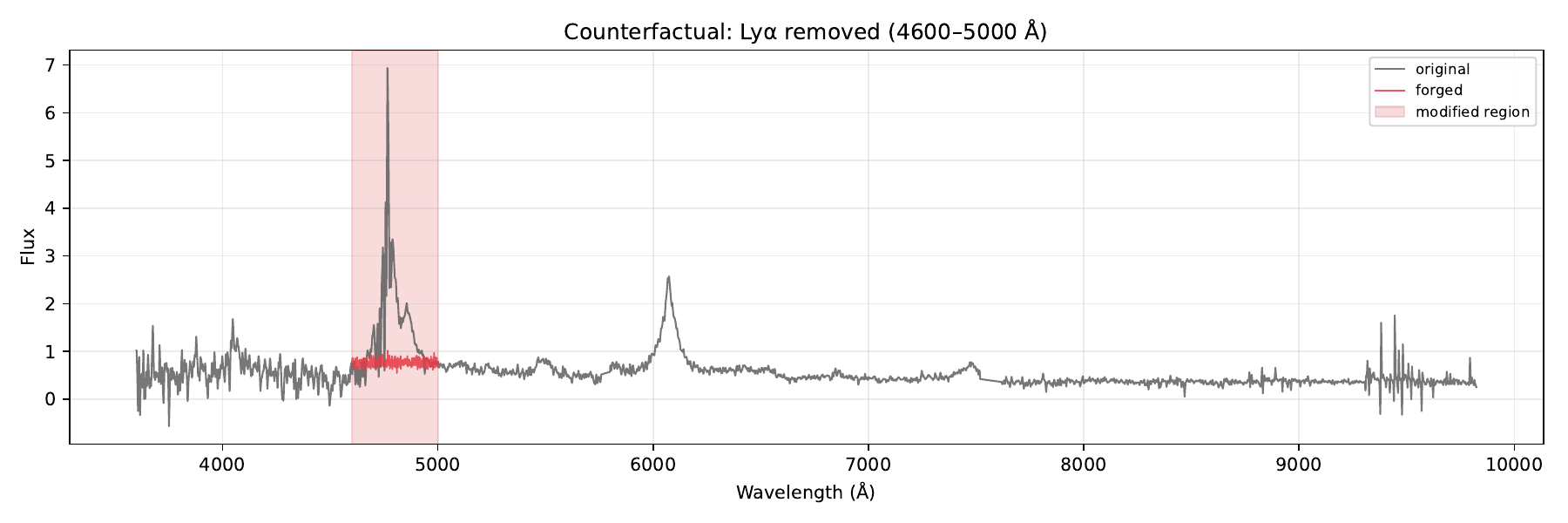}
    \includegraphics[width=0.99\linewidth]{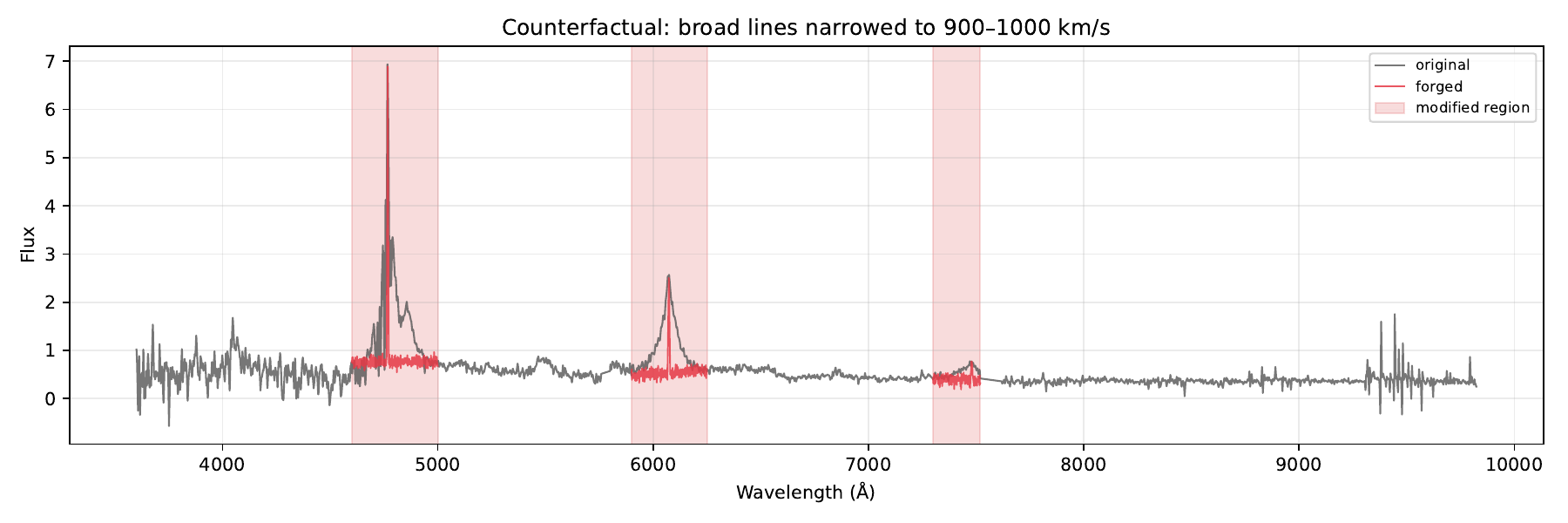}
    \includegraphics[width=0.99\linewidth]{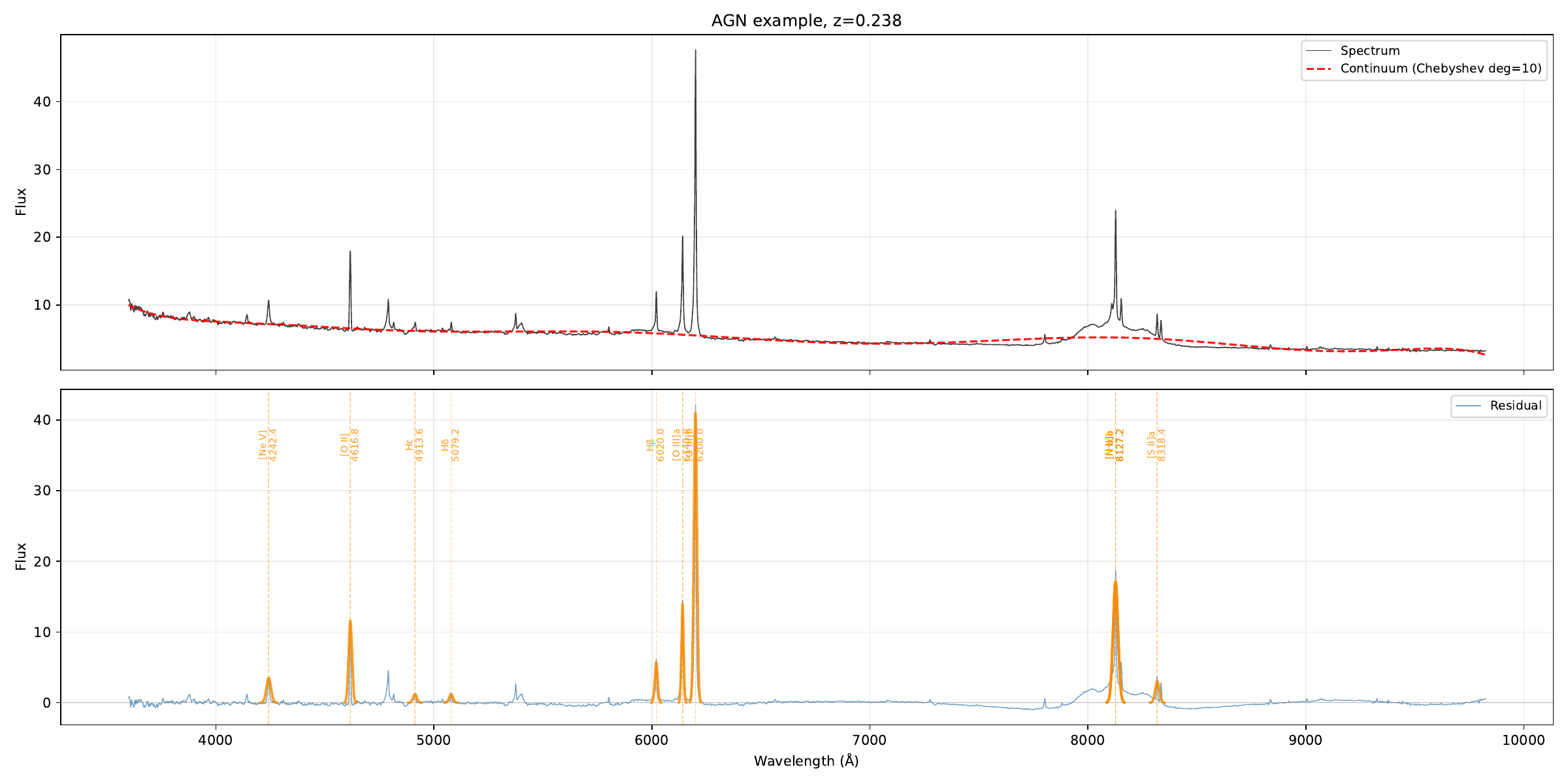}
\caption{\textbf{Evidence-based verification of physical interpretations.} \textbf{(a)} Counterfactual evidence-ablation tests. The spectrum is modified only in diagnostic regions to test whether the accepted interpretation depends on physically meaningful evidence. Removing the Ly$\alpha$ region or narrowing broad emission lines weakens the evidence required for a broad-line QSO interpretation, causing the verification audit to reject or defer the modified interpretation. This demonstrates that FORMA does not simply assign a label, but tests whether the physical evidence supporting that label remains valid. \textbf{(b)} Verification-supported recovery of a low-luminosity AGN interpretation. A standard galaxy interpretation can suppress the AGN nature of a host-dominated source, whereas expert visual inspection identifies a low-luminosity AGN candidate. FORMA supports the expert-VI interpretation by identifying high-ionization and AGN-relevant diagnostics, including [Ne~{\sc v}], strong [O~{\sc iii}] and C~{\sc iii}], and flags the object for audit or follow-up rather than treating it as a confirmed discovery.}
\label{fig3}
\end{figure*}

\clearpage

\bibliography{sn-bibliography}

\section*{Method}

\subsection*{Design rationale}

The architecture of FORMA is derived from structured interviews with DESI collaboration members and analysis of visual inspection protocol documentation. These investigations identified a reproducible expert workflow, illustrated in Fig.~\ref{fig1}a, comprising three stages. (1)~\textbf{Visual Perception}: experts parse the graphical representation of a spectrum into quantitative observables, detect spectral features, and form preliminary physical hypotheses by assessing spectrum morphology against a mental library of template spectra. (2)~\textbf{Hypothesis Inference}: they iteratively test candidate hypotheses by predicting the observed-frame positions of associated spectral lines and checking consistency across multiple line-based redshift estimates. For each candidate line, experts compute
\begin{equation}
z_i = \frac{\lambda_{{\rm obs},i}}{\lambda_{{\rm rest},i}} - 1,
\label{eq:redshift}
\end{equation}
and assess the agreement of all derived $z_i$ values. (3)~\textbf{Result Validation}: experts maintain a reflective posture throughout the workflow---they repeatedly scrutinise their feature detections, distinguish genuine spectral lines from noise, and verify the physical plausibility of the final interpretation. The DESI visual inspection interface operationalises this process: it enables an expert to step through successive Redrock template fits, toggling between them and the original spectrum to validate or challenge their own identifications. The expert is not passively accepting an automated output but conducting rapid, hypothesis-driven line matching, using model overlays as reference guides. This interplay of perceptual parsing, physically constrained hypothesis generation, and adversarial self-critique forms the cognitive template that FORMA formalises.

\subsection*{Architecture overview}

\begin{figure*}[p]
    \centering
    \includegraphics[width=0.99\textwidth]{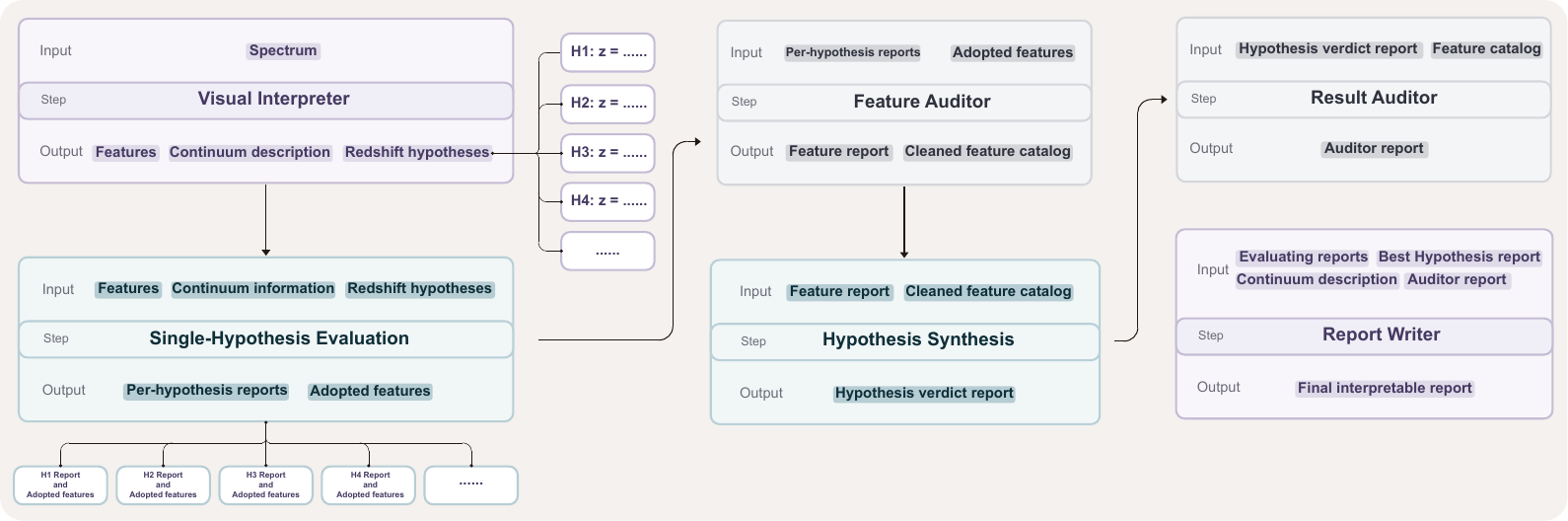}
    \label{fig4}
    \caption{The input and output of each step.
    }
    \label{fig:io of each step}
\end{figure*}

FORMA decomposes spectroscopic inference into four modules, three of which map directly to the stages of expert reasoning identified above, plus a synthesis and output module (Fig.~\ref{fig1}c). The modules are: (1)~\textbf{Visual Interpreter}, which reconstructs calibrated wavelength--flux arrays and generates redshift hypotheses through brute-force line matching; (2)~\textbf{Hypothesis Analyst}, which evaluates each hypothesis via parallel LLM agents equipped with domain-specific tools and synthesises a unified verdict; (3)~\textbf{Analysis Auditor}, which conducts two-stage adversarial review---first cross-verifying individual spectral features across hypotheses (Feature Auditor), then independently auditing the synthesis conclusion (Result Auditor); and (4)~\textbf{Report Writer}, which consolidates all upstream outputs into a structured report with a feature score on the 0--4 DESI VI scale and a credibility rating (HIGH/MEDIUM/LOW) from the Result Auditor. The four modules comprise six steps, each input and output are illustrated in Fig.~\ref{fig:io of each step}.

Perception, physical interpretation, numerical computation, and critical review each operate under distinct constraints that a monolithic model cannot simultaneously satisfy: a single LLM performing all steps is vulnerable to premature convergence on a plausible but incorrect interpretation, spurious consistency across logically independent constraints, and arithmetic errors in wavelength-to-redshift conversion. Isolating these functions into separate, auditable modules removes these failure modes at the architectural level. Numerical computation is strictly separated from LLM reasoning: wavelength-to-redshift conversion, spectrum access, line fitting, and redshift uncertainty propagation are implemented as deterministic functions and exposed via a tool-calling interface, with each module carrying only the tools required for its task. This separation ensures that quantitative reasoning is reproducible, auditable, and independent of the LLM's internal arithmetic.

\subsection*{Module 1, Step 1: Visual Interpreter}

The Visual Interpreter reconstructs a calibrated, quantified spectral representation from the raw input, producing the evidentiary foundation for all subsequent reasoning. The module extracts data from calibrated FITS files and produces a common internal representation comprising wavelength-calibrated flux arrays, per-pixel SNR estimates, fitted Chebyshev-polynomial continuum, and catalogues of spectral features.

\textbf{Spectrum reconstruction.}
For FITS input, the module loads spectral data directly from binary table HDUs. Taking DESI data as an example, B, R, and Z arm spectra are extracted, \texttt{SPECMASK}-flagged pixels are excluded, and inverse-variance or SNR arrays provide per-pixel noise information. Overlap regions between adjacent arms are masked to avoid stitching artifacts.

\textbf{Feature detection.}
Feature detection employs the Continuous Wavelet Transform (CWT) with a Mexican-hat (Ricker) wavelet applied at multiple scales via PyWavelets~\citep{Lee2019}. The algorithm detects emission peaks and absorption troughs, generating candidate feature catalogues with central wavelengths, amplitudes, FWHM (in \AA\ and km\,s$^{-1}$), and quality flags. Central wavelengths carry typical uncertainties of $10^0$--$10^1$\,\AA, propagating to redshift uncertainties at the $10^{-2}$--$10^{-3}$ level.

\textbf{Continuum fitting.}
Detected features are masked, and a Chebyshev polynomial continuum is fitted to the remaining flux points using specutils~\citep{nicholas_earl_2026_18473282}, with the polynomial degree selected automatically by balancing residual scatter, skewness, and model complexity. The fitted continuum is partitioned into approximately monotonic intervals, from which a structured natural-language description of the blue-to-red slope trend is generated to serve as a morphological reference for downstream classification.

\textbf{Hypothesis generation.}
The Visual Interpreter invokes the DESI \textsc{Redrock} template-fitting pipeline to generate redshift hypotheses. \textsc{Redrock} returns, for each of its top-$N$ fits, a redshift $z$, a spectral classification, and goodness-of-fit statistics ($\chi^2$, $\Delta\chi^2$). Of these, only the redshift is forwarded to downstream modules; the classification is determined autonomously by the Hypothesis Analyst and Analysis Auditor through physical line matching, independent of the template-based label. 

In fact, the hypotheses generation mechanism is not tied to any specific pipeline: template-fitting approaches are broadly applicable across spectroscopic surveys, and \textsc{Redrock} is used here as the baseline engine provided by DESI. Downstream reasoning modules consume hypotheses agnostic to their origin, so the outputs of any comparable redshift-fitting pipeline can be substituted directly.

\subsection*{Modules 2 \& 3: Hypothesis Analyst and Analysis Auditor}

The Hypothesis Analyst (Module~2) and Analysis Auditor (Module~3) are arranged as interleaved steps in a closed evaluation loop, mirroring the iterative, self-critical character of expert inspection. The workflow proceeds through four sequential steps:

\subsubsection*{Step 2: Single-hypothesis evaluation.}
Each of the $\sim 10$--$20$ candidate hypotheses generated by \textsc{Redrock} is assigned to an independent LLM agent equipped with three deterministic tools: single-Gaussian and double-Gaussian fitting for line-profile measurement around predicted positions, and redshift computation from observed and rest-frame wavelengths. The agent is designed to evaluate a single hypothesis against the pre-computed evidence, rather than conduct open-ended exploration.

Each agent receives a structured user prompt containing the hypothesis redshift with a broad verification window ($z \pm 0.1$), a Spectrum Summary (wavelength coverage, median SNR, masked regions), and a Predicted Lines table listing every rest-frame atomic line together with CWT-pre-detected features whose implied redshift falls within the verification window. A skill prompt, loaded as the system message, encodes the evaluation protocol: the agent first inspects the CWT features associated with each predicted line, adopting the best match by wavelength offset, multiscale CWT ridge persistence, and CWT SNR; only when no CWT detection exists does the agent invoke Gaussian fitting tool as a fallback. Each line is assigned a status of LIKELY, MARGINAL, NOT\_FOUND, or MASKED. The agent then selects a systemic redshift anchored on the lowest-ionisation LIKELY line, classifies the object by comparing the confirmed line inventory against type-specific diagnostic rules, and writes a structured adopted line catalogue and a natural-language report. All wavelength-to-redshift conversions are handled by the deterministic redshift tool, ensuring numerical reproducibility. The outputs are produced independently for each hypothesis, with no cross-hypothesis comparison at this step.

\subsubsection*{Step 3: Feature Auditor.}
Before hypotheses are compared, the Feature Auditor verifies the reality of individual claimed features across all hypotheses. It reads all per-hypothesis adopted line catalogs and identifies: (1)~lines claimed by one hypothesis but absent in others at similar observed wavelengths; (2)~known doublet pairs where only one component is claimed (orphan detection); (3)~doublet pairs with physically inconsistent amplitude ratios (Ca~K/H, [O~III], [N~II], [S~II]); and (4)~potential emission--absorption composite profiles, to help recognize the situation such as a broad Mg~II emission with a background galaxy Mg~II absorption on it.

An LLM agent equipped with raw-spectrum inspection and knowledge-base retrieval tools inspects each contested feature at its observed wavelength, assessing whether it is a genuine spectral line or a noise artifact. It checks [O~II] and Ly$\alpha$ forest morphology where applicable and verifies doublet separation and amplitude ratio consistency. Each feature receives a verdict of KEEP (physically real), REMOVE (noise or artifact), or FLAG (ambiguous, noted for downstream caution). The cleaned catalogs replace the originals for the synthesis step.

\subsubsection*{Step 4: Hypothesis synthesis.}
After the Feature Auditor has cross-verified individual features, a dedicated synthesis LLM agent performs cross-hypothesis comparison. The agent receives all per-hypothesis reports, the Feature Auditor's cleaned line catalogues, and tools for raw-spectrum inspection and knowledge-base retrieval, together with the Feature Auditor's pre-computed [O~II] morphology assessments. It builds unified line tables across hypotheses, reads raw spectrum regions to resolve conflicting claims, queries the knowledge base for classification diagnostics, assesses the relative evidential support for each hypothesis, and delivers a structured verdict containing: the best hypothesis (and optionally a runner-up), the adopted line pairs with supporting evidence, an intermediate confidence level (LOW/MEDIUM/HIGH), and remaining concerns. If no hypothesis meets the minimum plausibility threshold, the synthesis outputs a \texttt{Cannot confirm} verdict, triggering the downgraded inference procedure.

\subsubsection*{Step 5: Result Auditor.}
After synthesis, the Result Auditor performs an independent adversarial review of the synthesis verdict. Crucially, the auditor has no access to the synthesis agent's reasoning chain---it receives only the verdict, the adopted line catalogue, and the same tool set as the Feature Auditor (raw-spectrum inspection, knowledge-base retrieval, and morphological analysis) to construct its own reasoning from first principles. It systematically re-examines each key claim: re-reading the spectrum at suspicious line positions, querying the knowledge base to verify that the line identifications satisfy astrophysical constraints, checking for overlooked contradictions, and examining whether alternative interpretations were adequately considered. The output is a report containing: calibrated assessment---CONFIRM (no line revisions needed; classification physically consistent), NEEDS\_REVISION (line revisions or spectrum issues found that affect confidence), or UNCERTAIN (spectrum is noise-dominated or evidence is insufficient to decide)---together with a final credibility rating (LOW/MEDIUM/HIGH) and a summary of any remaining spectrum issues. 

\subsection*{Module 4, Step 6: Report Writer}

The Report Writer consolidates upstream outputs---spectrum metadata (wavelength coverage and SNR), continuum description, per-hypothesis cleaned line tables, Feature Auditor structured verdicts, synthesis verdict, and Result Auditor assessment---into a structured six-section final report: (1)~basic spectral information (wavelength coverage, continuum morphology, SNR, and edge-zone notes); (2)~hypothesis summary (a table of all evaluated hypotheses with line counts, anchor lines, and key strengths and weaknesses; a paragraph for the accepted hypothesis and one-sentence rejections for the excluded); (3)~synthesis and audit judgments (the Hypothesis Synthesis conclusion and Result Auditor's calibrated assessment presented side by side, with explicit notation of any disagreements); (4)~potential issues, organised into spectrum quality, line-identification uncertainties, physical consistency, and completeness categories; (5)~comprehensive assessment (final object type, recommended redshift with $\sigma_z$ uncertainty computed via a dedicated tool, confirmed line inventory, a feature score on the 0--4 scale, credibility rating, and a manual-review recommendation); (6)~a concise natural-language conclusion for non-specialist readers.

When the synthesis step returns no valid hypothesis, the Report Writer performs a downgraded inference: the strongest peak (by amplitude rank) drives a rule-based classification using its width class and continuum morphology, yielding a provisional object type but no redshift or line identifications. For instance, a broad emission peak (FWHM $> 2000$ km s$^{-1}$) on a blue continuum suggests a QSO, while narrow features on a red continuum favour a galaxy.

\subsection*{Feature score and credibility rating}

The final report carries two assessments of reliability. The \textbf{feature score} mimics DESI visual inspection quality (VI\_QUALITY), encoding the number and robustness of confirmed spectral features on a 0--4 scale \citep{Lan2023DESIVI,alexander23}:

\begin{description}
    \item[4] Confident classification---two or more secure features.
    \item[3] Probable classification---at least one secure feature with continuum support, or many weak features.
    \item[2] Possible classification---one strong emission feature, but its identity is uncertain.
    \item[1] Unlikely classification---one or more unidentified features.
    \item[0] No signal---nothing detected.
\end{description}

Because the feature score merely counts confirmed features, in practice we find that an incorrect analysis may still achieve a score of 4 by claiming at least two spurious features; the score alone cannot guarantee correctness (see Appendix.~\ref{app:feature-score}). The \textbf{FORMA credibility} rating addresses this limitation. Assigned by the Result Auditor (Step~5), it reflects how well the synthesis verdict withstands independent adversarial scrutiny: HIGH (no issues found; all key lines visually confirmed), MEDIUM (minor issues or limited line inventory), or LOW (major revisions needed or spectrum quality prevents confident assessment).

The credibility rating is assigned entirely by the LLM agent, without a detailed, hand-crafted scoring specification. As an experimental application of agent-based reasoning to spectroscopic classification, FORMA instead provides the agent with access to a knowledge base containing classification standards for each spectral type; the agent applies these criteria autonomously during adversarial review. Our results indicate that the agent produces well-calibrated credibility ratings from these knowledge-base standards alone. Empirically, the most common triggers for a LOW credibility downgrade are: line-width inconsistencies (a narrow detected feature matched to a broad reference line, or vice versa); missing key diagnostic lines (absent broad permitted lines within the observed wavelength range for QSO candidates, absent [O~II] or [O~III] for ELG candidates, or weak/absent Ca~K/H with an indistinct 4000 \AA\ break for BGS/LRG candidates); confirmed features clustered at the high-noise blue or red edges of the spectrograph arm; and key features coincident with bright airglow lines, particularly within the OH forest. Authoritative credibility scoring protocols tailored to the requirements of specific astronomical surveys remain a direction for future work.

\subsection*{Tools and knowledge base}

\textsc{FORMA} assigns all quantitative operations to a suite of deterministic Python functions, isolating numerical computation from language-model reasoning. When an agent needs to measure a spectral line, compute a redshift, or inspect a wavelength region, it invokes the corresponding tool through a function-calling interface; the tool executes deterministically and returns structured results. This ensures that no quantitative conclusion depends on the LLM's internal arithmetic.

The tool suite spans four functional categories. \textbf{Spectrum measurement} tools perform Gaussian and doublet fitting around predicted line positions, read raw flux arrays over specified wavelength intervals, and detect [O~II] unresolved-doublet slope-change signatures. \textbf{Numerical operations} handle redshift computation and uncertainty propagation via $\sigma_z = \sigma_\lambda / \lambda_{\rm rest}$. \textbf{Knowledge-base retrieval} enables agents to search a curated set of documents covering spectral line reference tables, classification diagnostics for each object type, ionisation priorities, and composite emission--absorption profile interpretation. \textbf{Structured output} tools write the report and line catalogues to output directory.

The knowledge base itself is designed to be read and applied by LLM agents. It encodes the astrophysical standards that human experts use during visual inspection: the diagnostic lines that define each object type, the ionisation sequences that must be satisfied, the criteria for interpreting blended emission--absorption profiles, and the conditions that render a classification physically untenable. Agents retrieve relevant passages by keyword search and apply the standards within their reasoning, guided by skill prompts that specify how each standard should be used at each step of the pipeline.

All wavelength-to-redshift conversions and tolerance-based line matching are pre-computed during Module~1 and supplied to downstream agents as authoritative input fields. Runtime tools are invoked primarily as a fallback---when a pre-computed feature requires verification against the raw spectrum, or when the agent must resolve an ambiguity that the static input fields cannot settle. This division of labour keeps the LLM's role focused on reasoning rather than arithmetic.

\subsection*{Dataset Composition}
\subsubsection*{DESI EDR Visual Inspection Data} 

We construct our evaluation dataset from the DESI Early Data Release (EDR) Visual Inspection (VI) \citep{Lan2023DESIVI,alexander23} catalogs, which contain VI information for deep co-added spectra from the first phase of Survey Validation (SV1). The VI catalogs comprise five files totaling approximately 22,000 spectra: Bright Galaxy Survey (BGS, 2,718 sources), Luminous Red Galaxies (LRG, 3,561), Emission Line Galaxies (ELG, 10,315), Quasars (QSO, 3,779), and a supplementary missed-QSO sample (1,717). Each entry provides a VI redshift (\texttt{VI\_Z}), a VI quality flag (\texttt{VI\_QUALITY}, 0--4), and a VI spectral type (\texttt{VI\_SPECTYPE}: \texttt{star}, \texttt{galaxy}, or \texttt{QSO}).

We build our dataset through the following procedure. (1) We select entries by VI spectral type across all five catalogues: QSOs wherever \texttt{VI\_SPECTYPE} = \texttt{QSO} (the VI-QSO population is distributed across the pipeline QSO, missed-QSO, ELG, LRG, and BGS catalogues); ELGs, LRGs, and BGS from their respective pipeline-named catalogues where \texttt{VI\_SPECTYPE} = \texttt{GALAXY}. (2) For each selected entry, we use \texttt{TARGETID}, \texttt{TILEID}, \texttt{RA}, and \texttt{DEC} to retrieve the corresponding coadded spectrum from the DESI FITS files. (3) From each spectrum's SCORES extension, we read the pre-computed per-arm values \texttt{MEDIAN\_COADD\_SNR\_B}, \texttt{MEDIAN\_COADD\_SNR\_R}, and \texttt{MEDIAN\_COADD\_SNR\_Z} and sum them to form a three-arm quality indicator. (4) Within each class, we rank spectra by this three-arm sum in descending order, stratify into three quality tiers at a 4:4:2 ratio (high / medium / low), and randomly sample 300 spectra while preserving the tier ratio. To avoid ambiguity from duplicate observations, we require unique TARGETID--TILEID pairs across the sample; this constraint reduces the BGS draw to 249 spectra and yields a final evaluation set of 1,149 spectra (300 QSO + 300 ELG + 300 LRG + 249 BGS).

\clearpage

\begin{appendices}

\section{Limitations of the feature score}\label{app:feature-score}

The feature score described above is a direct count of confirmed spectral features, designed to mimic the DESI visual inspection quality flag (VI\_QUALITY, 0--4). Before examining why this emulation fails, it is worth examining what VI\_QUALITY itself encodes, and what it was designed to do.

The DESI visual inspection programme was established with two stated goals: to construct catalogues of sources with redshift and source-type identifications verified by human inspectors, and to assess the performance of the \textsc{Redrock} pipeline by comparing its output against these verified identifications \citep{Lan2023DESIVI,alexander23}. The resulting VI catalogue serves as a ``truth table'' for quantifying the redshift recovery rate---the fraction of spectra with robust VI redshifts (VI\_QUALITY $\ge 2.5$) for which \textsc{Redrock} returns a matching solution. In this framework, VI\_QUALITY functions as a confidence-weighted pass/fail filter: spectra exceeding the threshold are treated as reliable truth-table entries; those below are excluded from pipeline performance calculations.

The VI\_QUALITY definition combines two separate judgments into a single integer: the inspecting expert's confidence that the adopted redshift is correct, and the number of spectral features identifiable at that redshift ($\ge 2$ features $\rightarrow$ score 4). During inspection, experts view the spectrum overlaid with the \textsc{Redrock} best-fit redshift-classification template, which marks the expected positions of spectral features at the proposed redshift, and experts can tune the template in wavelength space to produce their own redshift hypothesis. Notably, DESI does not rely on a single inspector: each spectrum is examined by multiple inspectors independently, and their reports are automatically merged into a final VI result only when three conditions are simultaneously satisfied---the redshifts differ by $\Delta z < 0.0033$ ($\sim 990$ km s$^{-1}$), the quality values differ by at most 1, and all inspectors assign the same spectral type \citep{Lan2023DESIVI}. The VI protocol also provides optional indications for inspectors to flag a bad redshift fit (R), a bad spectral-type fit (C), or a bad spectrum with artifacts such as cosmic-ray or skyline-subtraction residuals (S). Yet these safeguards—flagging, multi-inspector consensus, and quality merging—share a common failure mode: at low SNR, the Redrock template may anchor every inspector's perception, so an ambiguous fluctuation at the template-predicted wavelength appears to be a genuine feature. When the template fit looks adequate, no inspector has reason to invoke a flag or dissent from the consensus; the protocol's safeguards are bypassed not by negligence, but by design.

The underlying protocol makes an implicit assumption---that visual inspection can reliably distinguish genuine spectral features from noise and artifacts. For targets whose diagnostic features are bright and unambiguous, that assumption holds. For emission-line galaxies, it does not. ELG classification depends on narrow emission lines ([O~II], [O~III], H$\beta$) whose typical fluxes of $\sim 10 \times 10^{-17}$ erg s$^{-1}$ cm$^{-2}$ \AA$^{-1}$ lie at the same surface brightness as the OH forest ($\sim$6--10 $\times 10^{-17}$ erg s$^{-1}$ cm$^{-2}$ \AA$^{-1}$). In this regime, no inspector---human or algorithmic---can reliably tell a real [O~II] doublet from a noise fluctuation at a forest line. By contrast, QSO classification rests on broad permitted lines (Ly$\alpha$, C~{\sc iv}) reaching 15--40 $\times 10^{-17}$ erg s$^{-1}$ cm$^{-2}$ \AA$^{-1}$, which are far above the confusion floor.

\begin{figure}[h]
    \centering
    \includegraphics[width=0.95\textwidth]{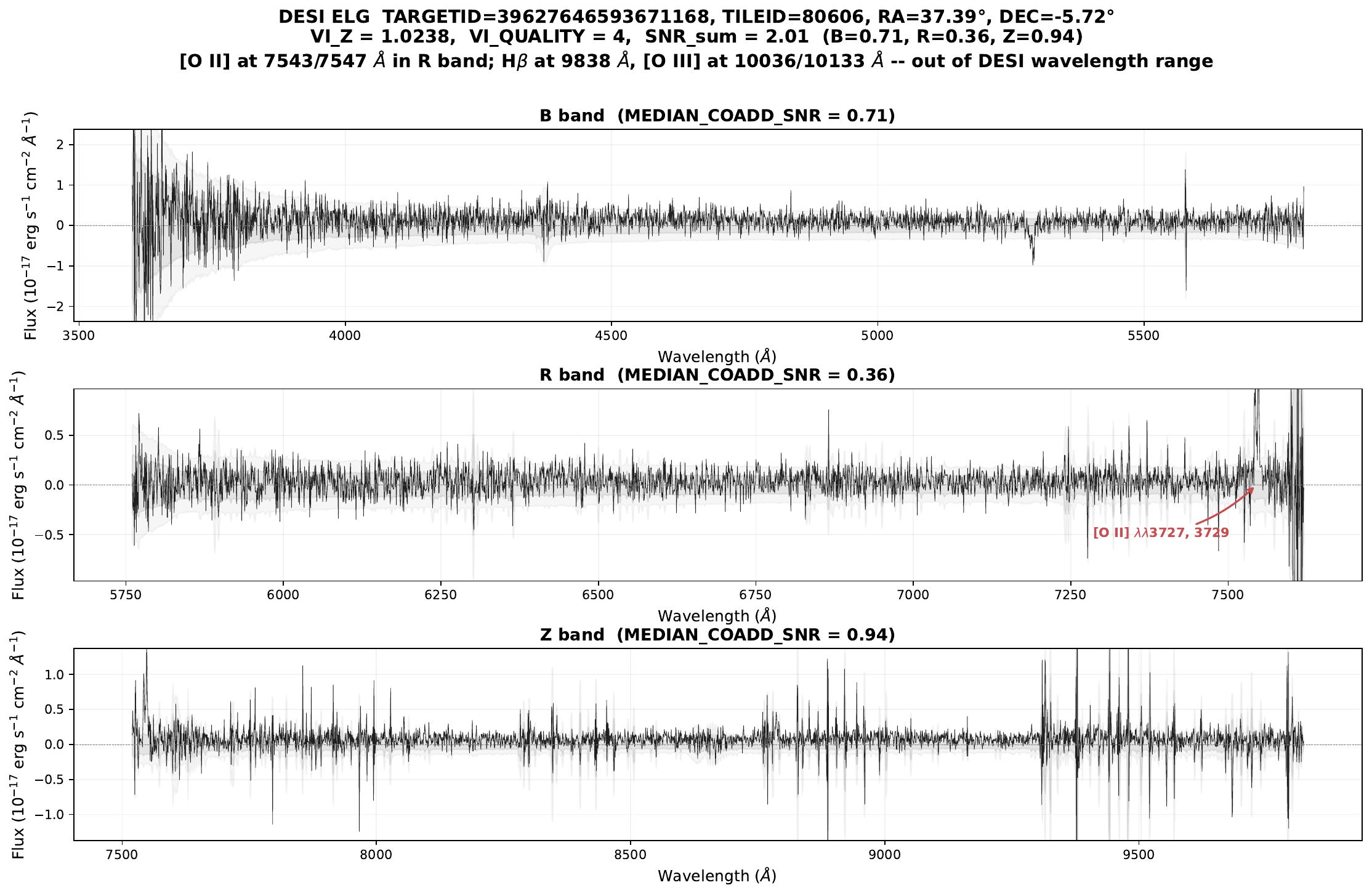} \\[6pt]
    \includegraphics[width=0.95\textwidth]{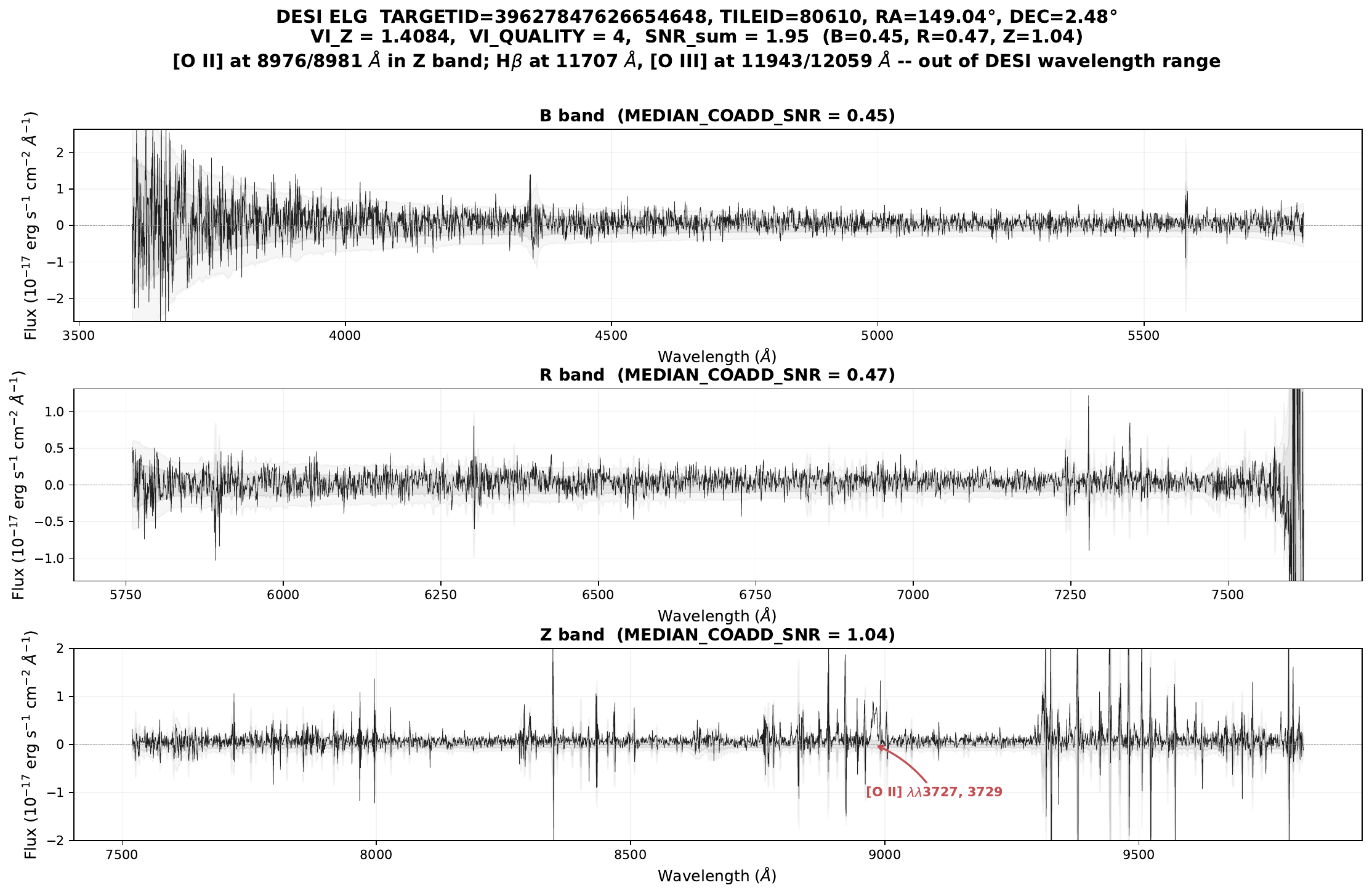}
    \caption{\textbf{Two DESI ELG spectra assigned VI\_QUALITY = 4.}
    \textbf{Top:} TARGETID 39627646593671168, $z_\mathrm{VI} = 1.0238$, three-arm \texttt{MEDIAN\_COADD\_SNR} sum = 2.0. The [O~II] doublet (arrow) falls in the R band where \texttt{MEDIAN\_COADD\_SNR\_R} = 0.36; H$\beta$ and [O~III] lie beyond the DESI wavelength range.
    \textbf{Bottom:} TARGETID 39627847626654648, $z_\mathrm{VI} = 1.4084$, three-arm sum = 1.95. [O~II] falls in the Z band (\texttt{MEDIAN\_COADD\_SNR\_Z} = 1.04); H$\beta$ and [O~III] are again inaccessible.
    In both cases, the only diagnostic feature available for classification is the [O~II] doublet at or below the $1\sigma$ noise level. Shaded bands indicate $\pm 1\sigma$ and $\pm 2\sigma$ noise envelopes.}
    \label{fig:elg-noise-examples}
\end{figure}

The consequence is that VI\_QUALITY loses semantic meaning in precisely the regime where it is most needed. In our evaluation set, ELG spectra assigned VI\_QUALITY = 4 have a median of 5.0 for the sum of \texttt{MEDIAN\_COADD\_SNR} (DESI pipeline output: median SNR per $\AA^{-1/2}$ in each arm) across the three DESI arms (per-arm median $\approx 1.7$), and 87\% fall below a three-arm sum of 10; a representative case (TARGETID 39627646593671168, Tile 80606, VI $z = 1.0238$) has a three-arm sum of 2.0, with \texttt{MEDIAN\_COADD\_SNR\_B} = 0.71, \texttt{MEDIAN\_COADD\_SNR\_R} = 0.36, and \texttt{MEDIAN\_COADD\_SNR\_Z} = 0.94. The [O~II] $\lambda\lambda$3727, 3729 doublet, which anchors the ELG classification at this redshift, falls in the R band at $\sim 7545$\,\AA---where \texttt{MEDIAN\_COADD\_SNR\_R} = 0.36, where the per-pixel noise envelope dominates over the typical signal; H$\beta$ and [O~III], the other ELG diagnostic lines required by the classification protocol, fall beyond the DESI wavelength coverage at this redshift (top panel of Fig.~\ref{fig:elg-noise-examples}). A second example at $z = 1.4084$ (bottom panel of Fig.~\ref{fig:elg-noise-examples}) tells the same story: [O~II] falls in the Z band where \texttt{MEDIAN\_COADD\_SNR\_Z} = 1.04, and H$\beta$ and [O~III] are again out of range.

In our evaluation set, QSO spectra with the same VI\_QUALITY = 4 have a median three-arm \texttt{MEDIAN\_COADD\_SNR} sum of 15.0 (per-arm median $\approx 5.0$), with only 30\% falling below a sum of 10. These are not the same quality: the QSO features are unambiguous detections, while the ELG features are claims whose supporting evidence is, at best, at the detection limit. A quality flag that assigns the same score to both is not measuring the physical reliability of the classification; it is measuring whether the inspecting expert, guided by a template, could identify at least two candidate features---a task whose outcome, at low SNR, is dominated by the expert's prior expectation rather than by the data. At these SNR levels, the expert's judgment is unavoidably anchored by the \textsc{Redrock} template solution: the template indicates where [O~II] should appear, and an ambiguous fluctuation at that location is indistinguishable from a genuine emission line. The protocol therefore cannot establish whether the expert is independently confirming a real spectral feature or merely ratifying a coincidence between the template and the noise. This limitation reflects a fundamental assumption of the VI protocol, not a shortcoming of the inspection team itself. 
%If a straight line with random noise were presented to an inspector alongside a template indicating that some of that noise happens to fall at the observed-frame wavelengths of [O~II] and [O~III], the protocol would likely return VI\_QUALITY = 4, because the protocol has no mechanism to assess whether a candidate feature is physically trustworthy once it passes the inspector's visual threshold. 
The protocol provides no explicit mechanism to distinguish whether an apparent feature corresponds to a genuine astrophysical signal or to a chance alignment between template expectations and noise fluctuations.
FORMA's credibility rating aims to address this gap: when the spectrum is noisy, the feature inventory is sparse, or line widths are inconsistent with the claimed classification, the Result Auditor is designed to downgrade confidence and issue NEEDS\_REVISION---a structured expression of uncertainty that VI\_QUALITY, by design, cannot provide.

The global distribution of VI\_QUALITY reflects this collapse of discriminating power. Across the full DESI VI catalogue of 22,090 spectra, 65\% receive VI\_QUALITY = 4 and 84\% receive $\ge 3$, leaving only 16\% below 3. A metric that places two-thirds of its population in the top category is not stratifying quality; it is functioning as a binary flag with a very permissive threshold, whose intermediate values convey little additional information.

\begin{figure}[h]
    \centering
    \includegraphics[width=0.99\textwidth]{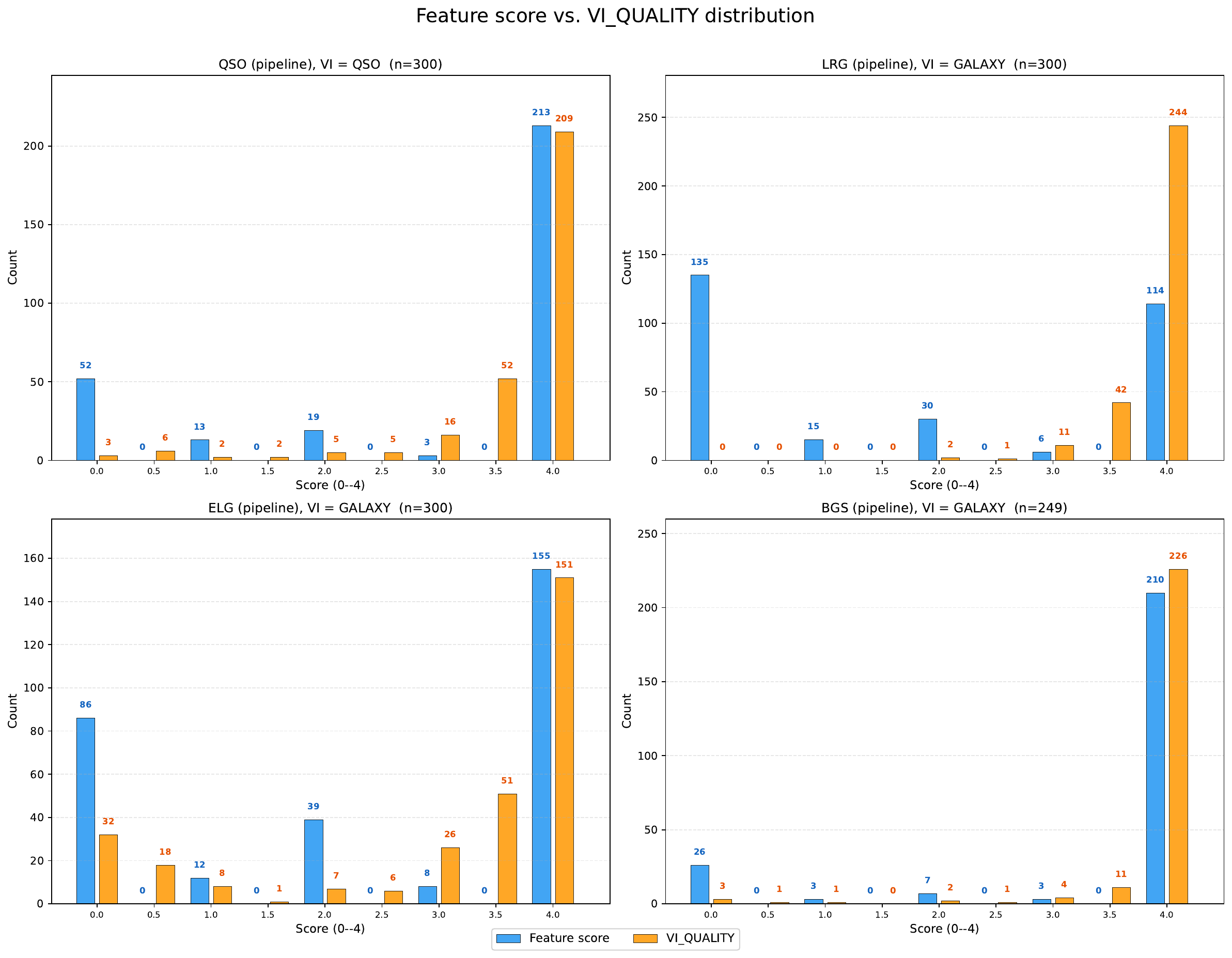}
    \caption{\textbf{Feature score vs.\ VI\_QUALITY distribution by pipeline class.} Blue bars show the feature score assigned by \textsc{FORMA}; orange bars show the expert VI\_QUALITY from the DESI VI catalogue. Both are binned at 0.5-unit intervals (0--4). The feature score spreads broadly for ELG and LRG despite their VI\_QUALITY being sharply peaked at 4, illustrating that line-counting conflates signal quality with classification confidence.}
    \label{fig:app-score-dist}
\end{figure}

Fig.~\ref{fig:app-score-dist} compares the feature score and VI\_QUALITY distributions side by side for each pipeline class. For QSOs, the two distributions are broadly aligned---both concentrate at score 4, consistent with the predominance of high-quality QSO spectra in the evaluation set. For ELGs and LRGs, however, a clear mismatch emerges: VI\_QUALITY is sharply peaked at 4, while the feature score spreads broadly across 0--4 with a large fraction at score 0. This is a direct consequence of the feature score's dependence on confirmed line count: ELG and LRG classifications rest on narrow emission lines that fall below a reliable detection threshold when \texttt{MEDIAN\_COADD\_SNR} is low; the downstream Feature Auditor and Result Auditor, which enforce line-width consistency, doublet ratios, and SNR checks, flag many of these marginal detections as untrustworthy---correctly excluding them from the confirmed line inventory. The feature score therefore counts fewer lines than the human expert claimed to see, not because the feature detector missed them, but because the adversarial audit layers refused to endorse detections that lack sufficient physical evidence. The resulting gap between feature score and VI\_QUALITY reflects a difference in evidentiary standards, not a failure of the automated pipeline.

\begin{figure}[h]
    \centering
    \includegraphics[width=0.48\textwidth]{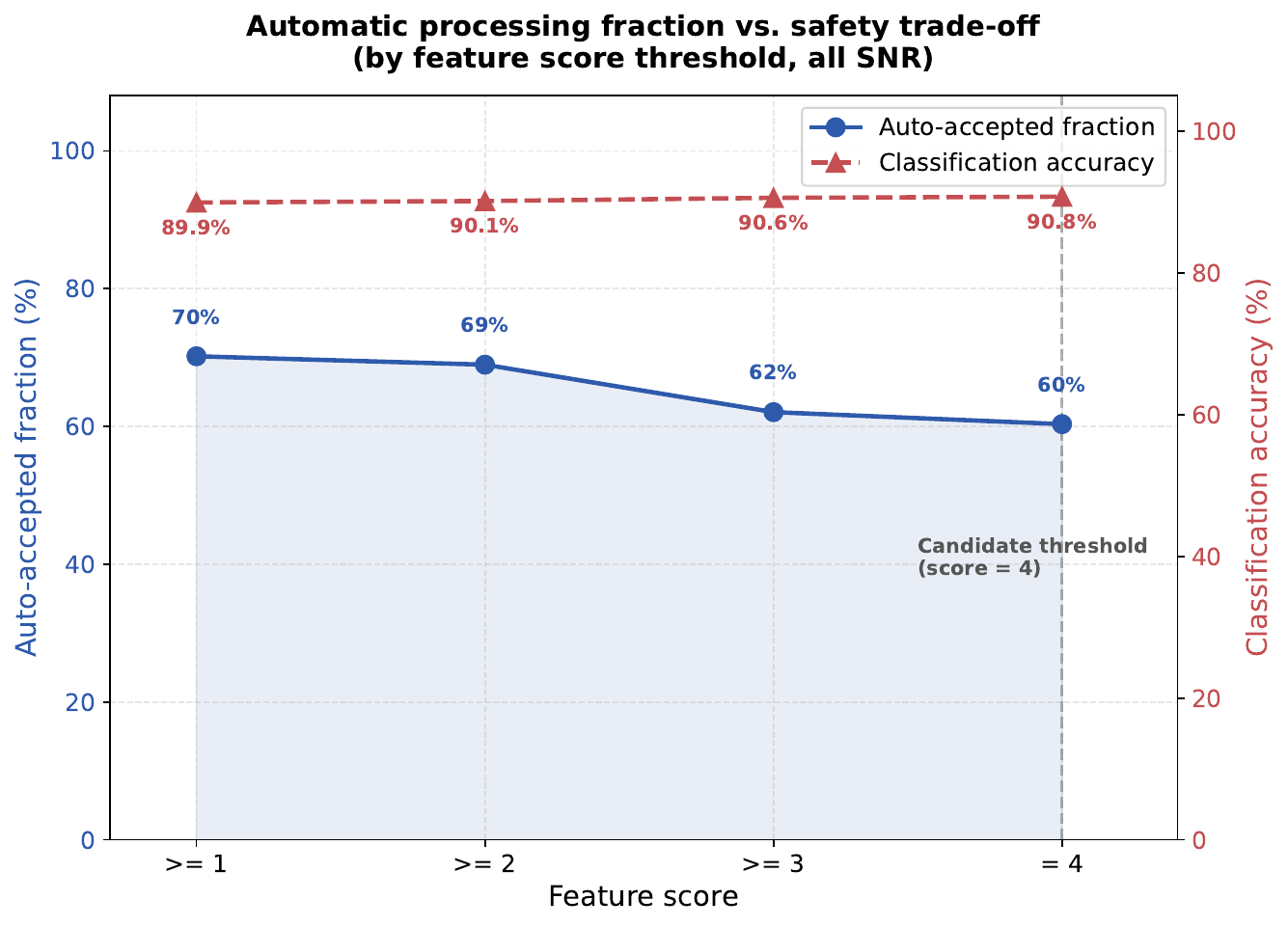}
    \includegraphics[width=0.48\textwidth]{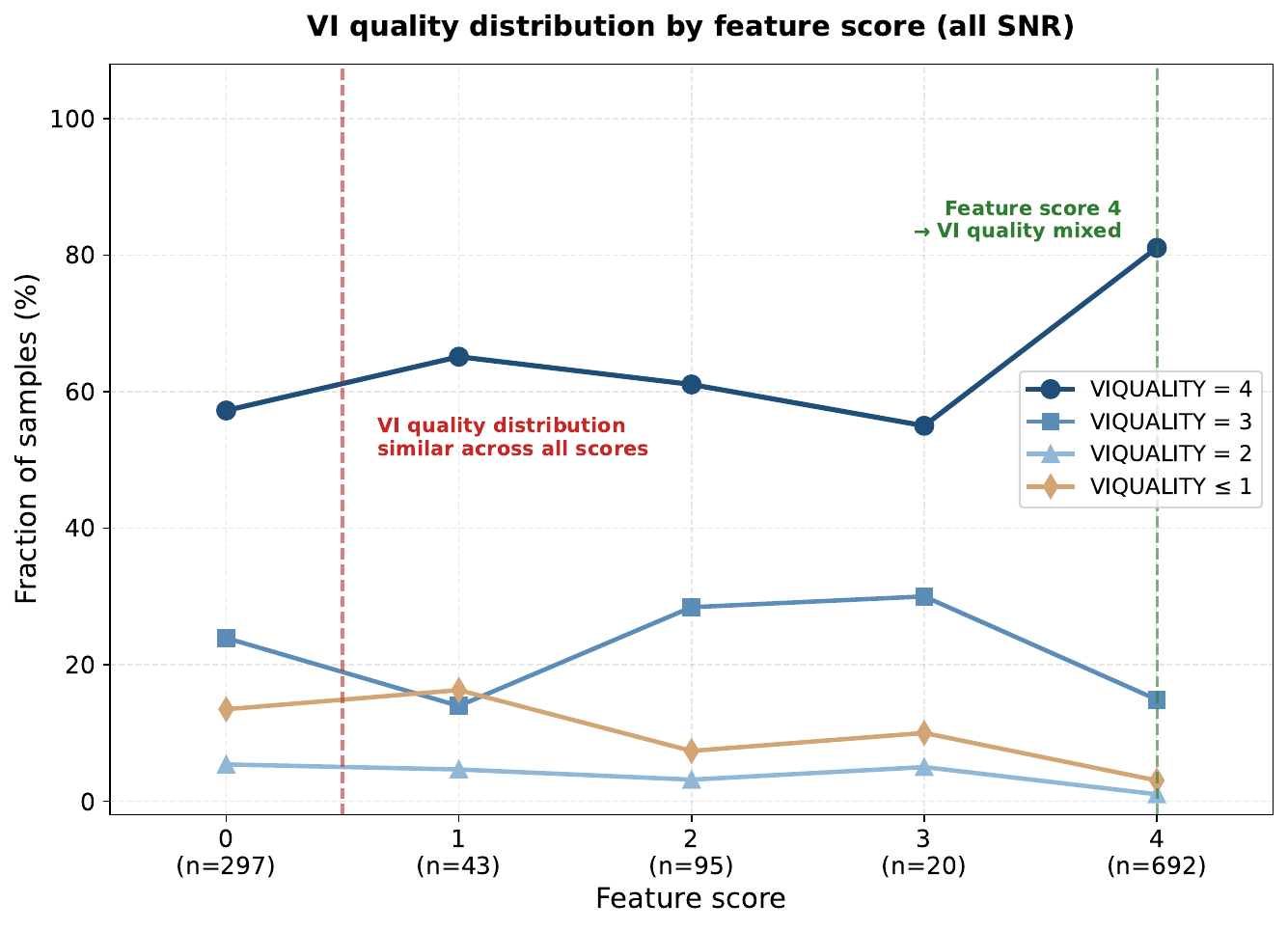}
    \caption{\textbf{Feature score does not recover the expert quality stratification.}
    (a) Automation--safety trade-off using the 0--4 feature score. Classification accuracy remains flat across thresholds, unlike the credibility rating where accuracy improves with stricter gating.
    (b) VI\_QUALITY distribution within each feature-score tier, using the full dataset without \texttt{MEDIAN\_COADD\_SNR} selection. The distribution of expert-assigned quality flags is nearly identical across feature-score bins, indicating that a simple feature count fails to reproduce the expert quality ranking.}
    \label{fig:app-score}
\end{figure}

Fig.~\ref{fig:app-score} provides a complementary view from two additional angles, using the same 1,149-spectrum evaluation set. Panel (a) shows the automation--safety trade-off using the 0--4 feature score: the classification accuracy remains approximately flat across thresholds ($\ge 1$ through $= 4$), in contrast to the credibility rating (HIGH/MEDIUM/LOW), where accuracy rises monotonically with stricter thresholds (see Fig.~\ref{fig2}a). Panel (b) shows the distribution of VI\_QUALITY within each feature-score tier. If the feature score successfully generalised the DESI VI convention, spectra with feature score 4 would be dominated by VI\_QUALITY = 4, and lower feature-score tiers would contain progressively lower VI\_QUALITY values. Instead, the VI\_QUALITY distribution is nearly identical across all feature-score bins---VI\_QUALITY = 4 accounts for a comparable fraction at every tier. Together, these panels confirm that a simple line-counting score conflates signal quality with classification confidence and does not recover the expert-assigned quality stratification, motivating the credibility rating as a complementary, reasoning-aware assessment.

\section{Counterfactual evidence-ablation tests}\label{app:counterfactual}

To test whether the Result Auditor (RA) rejects interpretations whose supporting evidence has been weakened or made physically inconsistent, we constructed counterfactual evidence-ablation experiments. Starting from an observed DESI spectrum for which \textsc{FORMA} returned a CONFIRM verdict at HIGH confidence for a broad-line QSO at $z=2.92$ (three confirmed broad emission lines: Ly$\alpha$, C~{\sc iv}, and C~{\sc iii}]), we constructed two counterfactuals, each involving a paired manipulation of the spectrum and the upstream catalogue that feeds the Result Auditor. In the first counterfactual, the Ly$\alpha$ emission region was removed from the spectrum and all entries associated with its observed wavelength ($\sim 4766$\,\AA) were deleted from the upstream line catalogue and hypothesis descriptions; the upstream pipeline thus receives a spectrum and catalogue in which Ly$\alpha$ lies within the observable wavelength range but yields no detection---a physically inconsistent scenario that nonetheless passes through to the RA as a valid QSO classification. In the second counterfactual, the broad emission lines were narrowed to FWHM $\approx 900$--$1000$ km s$^{-1}$ in the spectrum, and their upstream catalogue entries (at observed wavelengths $\sim 4766$, $\sim 6072$, and $\sim 7483$\,\AA) were correspondingly rewritten to describe them as narrow permitted lines. In both cases, the catalogue remains internally consistent with the altered spectrum, so no upstream module flags a contradiction; the RA must therefore detect the physical inconsistency from the spectrum and knowledge base alone. These modifications preserve a spectrum-like input but remove or alter the physical evidence required for a broad-line QSO interpretation. The modified spectra were passed through the full \textsc{FORMA} pipeline and the RA step was repeated 100 times independently to assess robustness against LLM sampling variability. The upstream modules (feature detection, hypothesis generation, single-hypothesis evaluation, Feature Auditor, and Hypothesis Synthesis) were held fixed; only the RA was re-run.

To assess whether the RA's behavior is driven by the knowledge base (KB) or by the spectrum alone, we repeated the experiment under two KB conditions for each counterfactual. In the \textbf{original} condition, the KB contains the standard astrophysical criteria (e.g., ``Ly$\alpha$ absent within the observable wavelength range is a fatal inconsistency''; ``broad emission lines are required for a Type-1 QSO classification''). In the \textbf{modified} condition, the specific criterion relevant to each counterfactual is removed from the KB, leaving the RA to assess the counterfactual spectrum without an explicit textual rule flagging the manipulated feature.

The RA nearly always rejects or defers the modified interpretations under both KB conditions (Table~\ref{tab:counterfactual}). Under the original KB, Ly$\alpha$ removal triggers NEEDS\_REVISION in 97 of 100 runs (61\% LOW confidence); the narrowed-line counterfactual is even more decisive, with all 100 runs returning NEEDS\_REVISION (92\% LOW). When the relevant KB criterion is removed, the rejection rate drops but remains strong: 86 of 100 runs still return NEEDS\_REVISION for the Ly$\alpha$ counterfactual, and 99 of 100 for the narrow-line case. The confidence distribution shifts toward MEDIUM under the modified KB (75\% and 23\%, respectively), indicating that the KB amplifies the RA's certainty but is not the sole driver of its judgment---the spectrum itself carries sufficient information for the RA to identify the physical inconsistency in the vast majority of runs. In both conditions, the RA has no access to knowledge of the counterfactual manipulation. These results demonstrate that the adversarial review layer combines KB-guided reasoning with direct spectral evidence to reject or defer physically inconsistent interpretations, and that neither component alone is fully responsible for the observed behavior.

\begin{table}[h]
    \centering
    \caption{\textbf{Result Auditor verdicts across 100 independent runs for each counterfactual condition and knowledge base (KB) variant.}}
    \label{tab:counterfactual}
    \begin{tabular}{@{}lcccc@{}}
    \toprule
    \textbf{} & \multicolumn{2}{c}{\textbf{Ly$\alpha$ removed}} & \multicolumn{2}{c}{\textbf{Lines narrowed}} \\
    \cmidrule(lr){2-3} \cmidrule(lr){4-5}
    & \textbf{Original KB} & \textbf{Modified KB} & \textbf{Original KB} & \textbf{Modified KB} \\
    \midrule
    NEEDS\_REVISION & 97 & 86 & 100 & 99 \\
    CONFIRM         & 3  & 14 & 0   & 1 \\
    Confidence LOW  & 61 & 25 & 92  & 77 \\
    Confidence MEDIUM & 39 & 75 & 8  & 23 \\
    \bottomrule
    \end{tabular}
\end{table}

\end{appendices}
\end{document}